**Interactions of SARS-CoV-2 spike protein and transient receptor potential (TRP) cation channels could explain smell, taste, and/or chemesthesis disorders**

Halim Maaroufi

Institut de biologie intégrative et des systèmes (IBIS). Université Laval. Quebec city, Quebec, Canada.

Halim.maaroufi@ibis.ulaval.ca

# IN BRIEF

SARS-CoV-2 S protein contains two ankyrin repeat binding motifs (S-ARBMs) that could interact with ankyrin repeat domains (ARDs) of transient receptor potential (TRP) channels. The latter play a role in olfaction, taste, and/or chemesthesis (OTC), suggesting that their dysfunction by S protein could explain OTC disorders in COVID-19 disease. Pharmacological manipulation of TRPs-ARDs could be used for prophylactic or treatments in SARS-CoV-2 infections.

# HIGHLIGHTS

- SARS-CoV-2 infection is associated to olfactory, taste, and/or chemesthesis (OTC) disorders (OTCD).
- Transient receptor potential (TRP) channels play a role in OTC.
- S protein contains ARBMs that could interact with ARDs of TRPs, thus inducing OTCD.
- TRPs ligands as possible preventive treatments against COVID-19.

# ABBREVIATIONS

SARS-CoV-2: severe acute respiratory syndrome coronavirus 2

COVID-19: Coronavirus disease-2019

S protein: Spike protein

RBD: Receptor binding domain

HR1: Heptad repeat 1

OTC: olfactory, taste and chemesthesis

OTCD: olfactory, taste and/or chemesthesis disorders

TRP: Transient receptor potential

TRPA1: Transient receptor potential Ankyrin 1

TRPC: Transient receptor potential canonical

TRPV: Transient receptor potential vanilloid

AR: Ankyrin repeat

ARD: Ankyrin repeat domain

TRP-ARD: Transient receptor potential-Ankyrin repeat domain

ARBM: Ankyrin repeat binding motif

S-ARBM: S protein-Ankyrin repeat binding motif

DARPins: Designed Ankyrin Repeat Proteins

mAbs: monoclonal antibodies

**Interactions of SARS-CoV-2 spike protein and transient receptor potential (TRP) cation channels could explain smell, taste, and/or chemesthesis disorders**

Halim Maaroufi

Institut de biologie intégrative et des systèmes (IBIS). Université Laval. Quebec city, Quebec, Canada.

Halim.maaroufi@ibis.ulaval.ca

**SUMMARY**

A significant subset of patients infected by SARS-CoV-2 presents olfactory, taste, and/or chemesthesis (OTC) disorders (OTCD). These patients recover rapidly, eliminating damage of sensory nerves. Discovering that S protein contains two ankyrin repeat binding motifs (S-ARBMs) and some TRP cation channels, implicated in OTC, have ankyrin repeat domains (TRPs-ARDs), I hypothesized that interaction of S-ARBMs and TRPs-ARDs could dysregulate the function of the latter and thus explains OTCD. Of note, some TRPs-ARDs are expressed in the olfactory epithelium, taste buds, trigeminal neurons in the oronasal cavity and vagal neurons in the trachea/lungs. Furthermore, this hypothesis is supported by studies that have shown: (i) respiratory viruses interact with TRPA1 and TRPV1 on sensory nerves and epithelial cells in the airways, (ii) the respiratory pathophysiology in COVID-19 patients is similar to lungs injuries produced by the sensitization of TRPV1 and TRPV4, (iii) resolvin D1 and D2 shown to reduce SARS-CoV-2-induced inflammation, directly inhibit TRPA1, TRPV1, TRPV3 and TRPV4, and (iv) liquiritin that inhibits TRPA1 and TRPV1 and protects against LPS-induced acute lung injury, is also reported as an inhibitor of SARS-CoV-2 infection *in vitro*. Herein, results of blind dockings of S-ARBMs, $^{408}$RQIAPG$^{413}$ (in RBD but distal from the ACE-2 binding region) and $^{905}$RFNGIG$^{910}$ (in HR1), into TRPA1, TRPV1 and TRPV4 suggest that S-ARBMs interact with ankyrin repeat 6 of TRPA1 near an active site, and ankyrin repeat 3-4 of TRPV1 near cysteine 258 supposed to be implicated in the formation of inter-subunits disulfide bond. These findings suggest that S-ARBMs affect TRPA1, TRPV1 and TRPV4 function by interfering with channel assembly and trafficking. After an experimental confirmation of these interactions, among possible preventive treatments against COVID-19, the use of DARPins, considered as potential alternatives to mAbs, to neutralise S-ARBMs and/or pharmacological manipulation (probably inhibition) of TRPs-ARDs to control or mitigate sustained pro-inflammatory response. In addition, it has been shown that TRPV1 is required for competent antibody responses to novel antigen. Probably, using agonists of TRPV1 as adjuvant in SARS-CoV-2 vaccination could enhance the quality and durability of immune response.

## INTRODUCTION

Since February 2020, chemosensory dysfunction in olfactory, taste and/or chemesthesis (OTC) was reported in a significant fraction of SARS-CoV-2 infected patients (von Bartheld et al., 2020). In the beginning, these first symptoms were considered as anecdotal, but today they are in the list of early disease symptoms. OTC give sensory warn that help to avoid inhalation, ingestion, or absorption of potentially harmful molecules. Particularly, chemesthesis has a protective function by sensing chemical irritants (Green et al. 1990; Jordt et al., 2004; Hata et al., 2012; Omar et al., 2017; Parma et al., 2020) and infectious agents, inducing their elimination from the airways rapidly by sneezing and coughing, and slowly by mucus secretion and inflammation (Green, 2012). The difference between OTC is often not clear because they give a unique sensation of flavor in the mouth. But, recent data suggest that taste and chemesthesis may be disturbed independently of smell in COVID-19 patients (Adamczyk et al., 2020; Lechien et al., 2020; Parma et al., 2020; Vaira et al., 2020). The somatosensory system (the trigeminal nerve) that conveys chemesthesis, is a separate sensory system with distinct peripheral and central neural mechanisms (Shepherd, 2006; Green, 2012). The trigeminal nerve may serve as a route for entry of pathogens into the brain, thus the RNA of coronavirus, mouse hepatitis virus strain JHM was detected in the trigeminal and olfactory nerves (Perlman et al., 1989). This route of invasion by SARS-CoV-2 could explain some neurological symptoms shown by COVID-19 patients, like for example, loss of facial sensation and headaches (Glezer et al., 2020). Recently, an observational study with more than two million persons revealed that olfactory and taste disorder is more predictive of COVID-19 than symptoms as fatigue, fever, or cough (Menni et al., 2020; Cooper et al., 2020; Sudre et al. 2020), often they are the only signs of the disease (Lechien et al., 2020; Hopkins et al., 2020). In addition, it has been reported that short time (main duration is 9 days) for a functional recovery (anosemia) could not be explained by neuron death, but probably by a support-cell mediated dysfunction of the olfactory epithelium (Heydel et al., 2013; von Bartheld et al., 2020). Of note, ethnic differences have been observed in OTCD, Caucasians had a 3-6 times higher prevalence of OTCD than East Asians (von Bartheld et al., 2020). This suggested a genetic specificity of SARS-CoV-2 interaction with proteins implicated in OTCD.

Ion channel proteins have an important role in the regulation of the function of innate immune system and participate to the pathogenesis of inflammatory/infectious lung diseases (Scheraga et al., 2020). The superfamily of the transient receptor potential (TRP) are weakly selective cation channels that are detected in plasma membrane of eukaryotes, from yeast to mammals (Venkatachalam and Montell, 2007). They are important components of $Ca^{2+}$ signaling pathways and regulate a wide range of physiological functions (Nilius B and Flockerzi, 2014). TRPs are principally $Ca^{2+}$ channels and were linked to viral infection. $Ca^{2+}$ cell entry is considered important for infection by different viruses, such as Sindbis virus, West Nile, HIV, filovirus, and arenavirus (Scherbik et al., 2010; Han et al., 2015; He et al., 2020). TRPV4 mediates $Ca^{2+}$ influx and nuclear accumulation of DDX3X in cells exposed to the Zika virus (Doñate-Macián et al., 2018). Respiratory syncytial virus and measles virus may interact directly and/or indirectly with TRPA1 and TRPV1 on sensory nerves and epithelial cells in the airways (Omar et al., 2017). And rhinovirus can infect neuronal cells and enhances TRPA1 and TRPV1 expression (Abdullah et al., 2014). Moreover, TRPs are considered as molecular sensors implicated in the detection of exogenous stimuli and endogenous molecules that signal tissue injury. For the defence of the respiratory airways, they induce airway constriction, sneezing and coughing, inflammation and mucus secretion (Xia et al., 2018; Emir et al., R.2017; Parker et al., 1998). Based on similarity in their primary sequences, TRPs are classified into seven subfamilies: TRPC (canonical, seven channels in human), TRPA (ankyrin, one), TRPV (vanilloid, six), TRPM (melastatin, eight), TRPML (mucolipin, three), TRPP (polycystin, three), and TRPN (nomp; absent in mammals) (Ramsey et al., 2006; Venkatachalam and Montell, 2007). They are homo or hetero-tetramers, each subunit forms six transmembrane segments (S1–S6) and a pore-loop between S5 and S6 that corresponds to a voltage-sensor-like domain and a pore domain. Each TRP channel subfamily is characterized by a unique cytoplasmic domain. The cytoplasmic N- terminal sequence of TRPA1, TRPC1, 3-7 and TRPV1-6 contains 16, 4 and 6 ankyrin repeat domains (ARDs), respectively. The 33 amino acids of ARDs form two alpha-helices linked by a beta-hairpin/loop (Gorina and Pavletich, 1996; Batchelor et al., 1998). In cell, the concave sides (ankyrin grooves) are facing the cell lipid bilayer. ARDs participate in diverse cellular functions by mediating specific protein–protein interactions and/or is an interaction site of endogenous or exogenous ligands (Barrick et al., 2008; Mosavi et al., 2004).

TRPA1, a unique gene in human, was first identified in human lung fibroblasts (Jaquemar et al., 1999). It was also detected in different cells, including human T cells and keratinocytes (Sahoo et al., 2019; Majhi et al., 2015; Assas et al., 2014; Omari et al., 2017). Moreover, it is expressed in the tongue of mammals and insects (Kim et al., 2010; Xiao et al., 2008). TRPA1 participates in inflammatory responses of the airways and is expressed at high levels in subjects with pathological conditions, such as chronic cough, asthma, rhinitis, and chronic obstructive pulmonary disease (Jaquemar et al., 1999; Mukhopadhyay et al., 2016). The work of Liu et al. (2020) suggests that an interaction between TRPV1/TRPA1 channels (TRPV1 is co-expressed with TRPA1 in nociceptive neurons) and the NF-κB pathway may be involved in increasing inflammation in the acute lung injury model. It has also been reported that TRPA1 is associated with inflammation and puritogen responses in dermatitis (Liu et al., 2013). The expression of TRPA1 and TRPV1 in nociceptive neurons and epithelial cells of the nasal cavity gives them a role of "warning system" against external and internal assaults (Story et al., 2003; Hata et al., 2012; Mukhopadhyay et al., 2016). Their activation by reactive electrophiles, elicits irritation, pain and inflammation (Zhang, 2015; Bautista et al., 2013; Zygmunt and Högestätt, 2014; Hasan et al., 2018). Interestingly, in vagal sensory neurons, interleukin-1 (IL-1) receptors are coexpressed with TRPA1. The latter is required to sense IL-1β, a central cytokine mediator of injury and inflammation. Moreover, TRPA1 in vagal neurons innervating the trachea and the lung, -deficient mice lack inflammatory reflex attenuation and fail to restrain cytokine release (Silverman et al., 2019).

TRPV1 is co-expressed with TRPA1 in nociceptive neurons (Story et al., 2003; Hata et al., 2012), and detected in almost all organs, including human T cells (Sahoo et al., 2019; Majhi et al., 2015), keratinocytes and lung fibroblasts (Assas et al., 2014; Omari et al., 2017). TRPV1 is also present in airway sensory fibers lining the trachea, bronchi, and alveoli, and the nasal mucosa (Groneberg et al., 2004; Seki et al., 2006; Watanabe et al., 2005b). The dysregulation of TRPV1 function has also been implicated in inflammation (Cao, 2020). Interestingly, in mice, aging reverses the role of the TRPV1 in systemic inflammation from anti-inflammatory to pro-inflammatory (Wanner et al., 2012).

TRPV4 is expressed in several non-neuronal cells like fibroblasts, smooth muscle, keratinocytes, vascular endothelial cells, macrophages, and in tracheal, bronchial, and alveolar epithelia (Vennekens et al., 2008; Alvarez et al., 2006; Jia et al., 2007; Palaniyandi et al., 2020). It participates in multiple physiological functions and pathological conditions, such as those related to epithelia, endothelium, osteoarticular tissues (Garcia-Elias et al., 2014; White et al., 2016; Nilius et al., 2013) and in innate immunity (Galindo-Villegas et al., 2016; Alpizar et al., 2017). Recently, it has been suggested that TRPV4 can participate in reducing viral infectivity in diseases such as dengue, Hepatitis C, and Zika (Doñate-Macián et al., 2018). Indeed, TRPV4 can regulate RNA metabolism dependent on DDX3X, a commonly expressed DEAD-box RNA-binding helicase that is typically hijacked by several RNA viruses (Yedavalli et al., 2004; Ariumi et al., 2007; Yedavalli et al., 2004). This suggests that TRPV4 could also play a role in SARS-CoV-2 infection (Doñate-Macián et al., 2018).

TRPC channels are ubiquitous in human tissues and regulate various cellular responses (Goodman and Schwarz, 2003; Louis et al., 2008). They are expressed in salivary glands, pulmonary and vascular smooth muscle cells (Beech et al., 2003). Indeed, TRPC3 and TRPC7 regulate respiratory rhythm (Ben-Mabrouk et al., 2010), whereas, TRPC6 has been shown important in the regulation of acute hypoxic pulmonary vasoconstriction and alveolar gas exchange (Weissmann et al., 2006). Targeting TRPC6 function may be used in therapeutics for the control of pulmonary hemodynamics/gas exchange. In addition, MxA, an interferon-induced GTPase, known to inhibit the multiplication of several RNA viruses, interacts with the ARD of TRPC1, -3, -4, -5, -6, and -7 (Lussier et al., 2005).

To prevent the development of severe COVID-19 form, it is necessary to understand the cellular basis of SARS-CoV-2 infection. Identifying sensory-neural mechanisms responsible of OTCD could help to find treatments to stop or mitigate COVID-19 development. Here, the potential molecular mechanism responsible of OTCD in SARS-CoV-2 infected patients is presented. Discovering that S protein contains two S-ARBMs and some TRPs cation channels, implicated in OTC, have ARDs, I hypothesized that the interaction of S-ARBMs and TRPs-ARDs could dysregulate the function of TRP channels and thus explains OTCD. Blind docking results suggested that probably S-ARBMs affect TRPs-ARDs function by interfering with channel assembly and trafficking.

## RESULTS AND DISCUSSION

### 1. Two ARBMs in S protein S1 and S2 subunits

The eukaryotic linear motif (ELM) resource (http://elm.eu.org/) revealed in SARS-CoV-2 S protein two short linear motifs (SLiMs) known as ARBMs (R-x-x-[PGAV][DEIP]-G) (Guettler et al., 2011), $^{408}$RQIAPG$^{413}$ (in RBD but distal from the ACE-2 binding region) and $^{905}$RFNGIG$^{910}$ (in HR1) (Fig. 1). ARBMs interact with proteins containing ARDs. Of note, in ELM, ARBMs are called tankyrase binding motif because they interact with the ARD region of Tankyrase-1 and -2 (Table 1). Importantly, $^{408}$RQIAPG$^{413}$ and $^{905}$RFNGIG$^{910}$ are localized in two important regions of S protein implicated in host cell attachment (RBD) and membrane fusion (HR1), respectively (Xia et al., 2020). In addition, ScanProsite tool (https://prosite.expasy.org/scanprosite/) was used to verify if ARBMs are present elsewhere in SARS-CoV-2, SARS-CoV and MERS-CoV proteomes. Indeed, they are only present in S proteins of SARS-CoV-2 and SARS-CoV, but absent in MERS-CoV (Fig. 1A, B). Interestingly, figure 1C shows that SARS-CoV-2, SARS-CoV, bat RaTG13 and pangolin GX-P5L S proteins with ARBMs in both S1 and S2 subunits are phylogenetically close. Of note, S2 subunit of the other beta-coronaviruses has an ARBM-like where Isoleucine in position five (R-x-x-[PGAV][DE**I**P]-G) is replaced by a similar physicochemical hydrophobic amino acid residue, Leucine or Valine (Fig. 1B).

ARBMs are present in S protein of SARS-CoV-2 and SARS-CoV (Fig. 1A, B), and their interaction with TRPs-ARDs could explain OTCD. Intriguingly, only one single case of anosmia was reported during the SARS-CoV pandemic (Hwang, 2006), against almost millions of cases in SARS-CoV-2 (von Bartheld et al., 2020). In the beginning of the SARS-CoV-2 pandemic, loss of olfactory and taste were considered as anecdotal. May be in the SARS-CoV pandemic, these first symptoms were also considered as anecdotal due to infection cases were not large (little more than 8,000) (Peiris et al., 2003), against millions of cases in SARS-CoV-2, to reveal clearly that loss of olfactory and/or taste are linked to SARS-CoV infection. In addition, it has been observed that Caucasians had a 3-6 times higher prevalence of OTCD than East Asians (von Bartheld et al., 2020). The SARS-CoV pandemic was for major part localized in China. For MERS-CoV, none case of olfactory and/or taste loss was reported. This is in accordance with the absence of ARBMs in MERS-CoV S protein. Finally, as a note, it has been reported that abnormalities of olfaction and taste may be major factors in the anorexia of acute viral hepatitis (Henkin and Smith, 1971).

## 2. $^{408}$RQIAPG$^{413}$ motif is in a hot region

Scanning SARS-CoV-2 S protein sequence with DisEMBL software (Linding et al., 2003) revealed in the RBD a hot disordered loop, $^{404}$RGDEVRQIAPGQTGKIA$^{419}$, that contains four motifs: $^{408}$**R**QI**AP**G$^{413}$ motif, reversed consensus pattern (TQ) phosphorylation (reading from C- to N-terminus) (Torshin, 2000), phosphothreonine motif (T-x-x-I) and RGD motif (Table 1).

### 2.1. Reversed consensus pattern (TQ) phosphorylation

The Thr415 of the reversed consensus pattern (TQ) (reading from C- to N-terminus) in $^{404}$RGDEVRQIAPGQTGKIA$^{419}$, could be phosphorylated by phosphatidylinositol 3- and/or 4-kinase (PIKKs). The Gln414 (Q) beside to the target Ser/Thr is critical for the substrate recognition. PIKKs are Ser/Thr atypical kinases that are found only in eukaryotes (Imseng et al., 2108; Angira et al., 2020). Phosphatidylinositol-3 kinase (PI3K) and PI4K are responsible for the production of phosphoinositides that are important in cell signaling. Interestingly, Yang et al. (2012) showed that phosphatidylinositol 4-kinase IIIβ (PI4KB) is required for cellular entry by pseudoviruses bearing the SARS-CoV S protein. This cell entry is highly inhibited by knockdown of PI4KB. They also demonstrated that PI4KB does not affect virus entry at the SARS-CoV S-ACE2 interface. Furthermore, Liu et al. (2005) demonstrated that PI4K that synthesis phosphatidylinositol 4,5-biphosphate (PIP2) is required for full recovery of TRPV1 from desensitization. It is known that the activation of TRPV1 leads to both pro-inflammatory and anti-inflammatory responses. The anti-inflammatory effect most likely results from TRPV1 channel desensitization (Liu et al., 2020). These above cited results suggest that PI4K is implicated in cellular entry of SARS-CoV by the synthesis of PIP2 (cell signaling). Indeed, it has been reported that PIP2 sensitizes TRPV1 (Liu et al., 2005; Stein et al., 2006). Thus, it is possible that sensitized TRPV1 permits virus cell entry and consequently produces pro-inflammatory effect in SARS-CoV/CoV-2 patients. Probably, using TRPV1 antagonists could inhibit or mitigate inflammation.

About $^{905}$RFNGIG$^{910}$, its C-terminal amino acids overlap with amino acids sequence $^{909}$IGV**TQ**N**V**L$^{916}$ that contains two motifs (phosphothreonine motif (TQ) and (T-x-x-I)) which Thr912 could be phosphorylated by PIKK family members, as described above for $^{408}$RQIAPG$^{413}$ motif, and CaM-II kinase, respectively.

## 2.2. Phosphothreonine motif ($^{415}$T-x-x-I$^{418}$) binding

As described above Thr415 is present in two patterns. When phosphorylated in the pattern ($^{415}$T-x-x-I$^{418}$), it binds a subset of forkhead associated (FHA) domains. For example, Herpes Simplex Virus type-1 (HSV-1) mimics a host cellular phosphosite in its E3 ubiquitin ligase (ICP0) to bind the host DNA damage response E3 ligase RNF8 via the RNF8 FHA domain. Thus, phosphorylation of ICP0 recruits RNF8 for degradation and thereby promotes viral multiplication (Chaurushiya et al., 2012). May be SARS-CoV-2 uses phosphorylated Thr415 to hijack cellular functions by binding RNF8 FHA domain or other proteins with FHA domain.

## 2.3. RGD motif

The RGD motif is the minimal peptide sequence used by many human viruses to bind proteins of the integrin family (Hussein et al., 2015). For example, RGD motif integrin-binding is essential for human Adenovirus type 2/5 (Wickham et al., 1994), Rotavirus (RV) (Zárate et al., 2004) and Kaposi's sarcoma-associated virus (HHV-8) (Hussein et al., 2015). Binding to integrin may play a supplemental role to ACE2 binding, like facilitating endocytosis by signaling through the integrin family. Thus, RGD promotes infection by binding integrin heterodimers formed by α and β subunits (Hussein et al., 2015), activating transducing pathways involving phosphatidylinositol-3 kinase (PI3K). This establishes a link with reversed consensus pattern (TQ) phosphorylation (paragraph 2.1) that Yang et al. (2012) showed that the kinase phosphorylating this motif is required for cellular entry by pseudoviruses bearing the SARS-CoV S protein.

Finally, to show the importance of phosphorylation associated to ARD in the entry of viruses into host cell, Than et al. (2016) reported that HCV NS5A protein interacts with ANKRD1 (by its C-terminal ARD) and Pim kinase. Both Pim kinase and ANKRD1 are involved in HCV entry step but not in cell attachment step.

## 3. Interactions of $^{408}$RQIAPG$^{413}$ and $^{905}$RFNGIG$^{910}$ with human ARDs of TRPA1, TRPV1 and TRPV4

SARS-CoV-2 is associated to significant cases of olfactory, taste, and/or chemesthesis disorders (OTCD) (von Bartheld et al., 2020). Because SARS-CoV-2 S protein contains two ARBMs (Fig. 1A, B) and some TRPs (TRPA1, TRPC1, TRPC3-7, and TRPV1-6) cation channels, implicated in OTC, have TRPs-ARDs, I hypothesized that the potential interaction of S-ARBMs and TRPs-ARDs could be responsible of OTCD. Some TRPs-ARDs are expressed in the olfactory epithelium, taste buds, trigeminal neurons in the oronasal cavity and vagal neurons in the trachea/lungs (Table 2). Furthermore, this hypothesis is supported by works that have shown that (i) respiratory viruses interact with TRPA1 and TRPV1 on sensory nerves and epithelial cells in the airways (Omar et al., 2017), (ii) the respiratory pathophysiology in COVID-19 patients is similar to lungs injuries produced by the sensitization of TRPV1 (Nahama et al., 2020) and TRPV4 (Kuebler et al., 2020) and (iii) liquiritin (one of the major flavonoids in *Glycyrrhiza uralensis*) that inhibits TRPA1 and TRPV1 and protects against LPS-induced acute lung injury (Liu et al., 2020), is also reported as an inhibitor of SARS-CoV-2 infection *in vitro* (Zhu et al., 2020).

In order to interact with ARDs, $^{408}$RQIAPG$^{413}$ and $^{905}$RFNGIG$^{910}$ must be exposed at the surface of S protein. Indeed, these motifs are exposed in the surface of RBD and HR1 (Fig. 3B). To compare the interactions of $^{408}$RQIAPG$^{413}$ and $^{905}$RFNGIG$^{910}$ with ARD region of human TRPA1, TRPV1 and TRPV4 and other co-crystallised complex of ARBM-ARD, blind docking was performed with Frodock (Garzon et al., 2009) and AutoDock vina (Trott and Olson, 2010) softwares. Before molecular docking, the reliability of these softwares was validated by re-docking the subunits of the complex ACE2-RBD (PDB id: 6m0j) and the complex of ARD of Tankyrase2-peptide SH3BP2 (PDB id: 3twr). Indeed, Frodock and Autodock Vina were able to produce a similar docking pose for each control protein with respect to its biological conformation in the co-crystallised protein-protein complex. Figure 3A shows that important amino acid residues R408 and G413 of $^{408}$RQIAPG$^{413}$ in RBD interact with ankyrin repeat 6 (AR 6) of TRPA1. AR 6 is a hot region in TRPA1 channel function. Indeed, it has been reported that the inhibition of hTRPA1 channel gating is possible via an AR 6 interaction (Tseng et al., 2018). And in mouse TRPA1, all mutations that make TRPA1 heat activated are located in AR 6 (Jabba et al., 2014). It has also been reported that ARD can bind lipids (Kim et al., 2014). Recently, Toelzer et al. (2020) observed that the anchor for the headgroup carboxyl of linoleic acid (LA) is provided by Arg408 and Gln409 of $^{408}$RQIAPG$^{413}$ from the adjacent RBD in the trimer. It is tempting to speculate that LA in binding pocket of S protein could modify the TRP channels activity through the interaction $^{408}$RQIAPG$^{413}$-ARDs. The importance of lipids in the activity of TRPs has been demonstrated in several studies. Resolvins RvD1 and RvD2 that are endogenous lipid mediators with pro-resolving and anti-inflammatory functions (Serhan et al., 2002; Serhan et al., 2008) inhibit directly TRPA1, TRPV1, TRPV3 and TRPV4 (Dhakal et al., 2019). Interestingly, RvD1 and RvD2 are recently shown to reduce SARS-CoV-2-induced inflammation (Recchiuti et al., 2020). In addition, LA stimulates $Ca^{2+}$ increase in pancreatic islet beta-cells through extracellular calcium influx via TRP channels (Wang et al., 2010). Oxidized LA metabolites activate TRPV1 (Patwardhan et al., 2009) and polyunsaturated fatty acids can sensitize, activate or inhibit vertebrate TRP channels including TRPV1 (Matta et al., 2007) and TRPV3 (Hu et al., 2006). Moreover, it had already been shown that the TRPV4 ARD interacts with the inositol head group of PI(4,5)P2, which negatively regulates the TRPV4 channel activity (Takahashi et al., 2014). PI(4,5)P2 was located beside the AR 4 and interacts with the residues from AR 3–5 by its phosphate groups.

The same result like RBD-TRPA1, shown above, was obtained with the docking of HR1 into ARD of TRPA1 (Fig. 3B). To confirm the results of TRPA1-RBD and TRPA1-HR1 complexes, $^{408}$RQIAPG$^{413}$ and $^{905}$RFNGIG$^{910}$ peptides were used in blind docking. The docked peptides were localized in the same region where were localized RBD and HR1 in ARD of TRPA1 (Fig. 3C). Structural alignment of docked $^{408}$RQIAPG$^{413}$ and $^{905}$RFNGIG$^{910}$ peptides with crystallised Tankyrase-2 AR-human SH3BP2 peptide (PDB id: 3twr) showed that these complexes were superposed (Fig. 3C). This suggested the reliability of the blind docking. Of note, positively charged Arg408 and Arg905 of S-ARBMs interact with negatively charged region of electrostatic potential surface representation of the ARD region of TRPA1 (Fig. 3D).

Docking results of TRPV1 ARD with RBD and HR1 are represented in Figure 4. The latter shows that RBD and HR1 interact with AR 3-4 of TRPV1 near cysteine 258 supposed to be implicated in the formation of inter-subunits disulfide bond (Tanaka et al., 2020). Contrary to TRPA1 and $^{408}$RQIAPG$^{413}$ and $^{905}$RFNGIG$^{910}$ complexes that are superposed with the same orientation (N- to C-terminal), TRPV1-peptides complexes have opposed orientation (Fig. 4D).

The results of blind docking of S-ARBMs, $^{408}$RQIAPG$^{413}$ and $^{905}$RFNGIG$^{910}$, into TRPA1, TRPV1 and TRPV4 (result not shown) suggest that S-ARBMs interact with AR 6 of TRPA1 near an active site, and AR 3-4 of TRPV1 (near cysteine 258) and TRPV4 (near cysteine 294). Cysteine 258 of TRPV1 is supposed to be implicated in the formation of inter-subunits disulfide bond (Tanaka et al., 2020). These findings suggest that S-ARBMs affect TRPA1, TRPV1 and TRPV4 function by interfering with channel assembly and trafficking. TRP channels transduce signals that help to avoid inhalation, ingestion, or absorption of potentially harmful molecules. And induce their elimination from the nose, oral cavity and airways rapidly by reflex responses such as sneezing and coughing (Green, 2012). It is tempting to speculate that SARS-CoV-2 interferes with this first line of the chemosensory defense to increase its ability to infect human.

Bibliographic search for the possible interactions of viruses and host proteins with ARD, revealed some results. Drappier et al. (2018) showed that when Theiler's murine encephalomyelitis virus (TMEV) L* protein binds to the ARD of RNase L, it inhibits 2'-5' oligoadenylates binding, and thus preventing the dimerization and oligomerization of RNase L. The latter is the effector enzyme of the OAS/RNase L system, interferon-induced antiviral pathways. Active RNase L (oligomeric form) degrades RNA of the infected cell and viruses, thus stopping virus spread. Moreover, in Hepatitis C virus (HCV), NS5A protein (nonstructural 5A) interacts with ANKRD1 (ankyrin repeat domain 1) protein by the ARD in its C-terminal region, thus participating in the entry step but not cell attachment step during HCV infection (Than et al., 2016). Li et al. (2020) reported that Influenza A virus (IAV) PA-X protein (a ribonuclease) interacts with the N-terminal ARD of ANKRD17. The latter is a positive regulator of inflammatory responses. This interaction attenuates the overactivation of the innate immune response to infection in host cells. Finally, Epstein-Barr virus (EBV) is a herpesvirus known in human to establish lifelong latent infections. Its protein EBNA1 (Epstein-Barr nuclear antigen 1) has two ARBMs in its N-terminal domain that interact with the ARD of tankyrase-1 and 2. This interaction downregulates OriP replication and plasmid maintenance (Deng et al., 2005).

## *4. Dusquetide, an anti-inflammatory peptide, presents similarity with $^{408}$RQIAPG$^{413}$*

In the aim to find small molecules containing substituents that topologically and structurally mimic R-x-x-[PGAV][DEIP]-G motif, sketch of 3D structure of $^{408}$RQIAPG$^{413}$ and $^{905}$RFNGIG$^{910}$ are generated and compared with molecules in PubChem (https://pubchem.ncbi.nlm.nih.gov/). A peptide called dusquetide with five amino acid residues (*RIVPA*) was identified. This peptide modulates the innate immune response to the inflammation caused by cell damage. It acts by binding to p62, a key adaptor protein that functions downstream to the key sensing receptors (e.g., toll-like receptors) that trigger innate immune activation (Yu et al., 2009). Dusquetide has been shown to significantly reduce IL-6 (North et al., 2016), which plays an important role in cytokine storm syndrome. It also modulates the cellular signaling from a pro-inflammatory to an anti-inflammatory response (North et al., 2016; Scott et al., 2007; Yu et al., 2009).

## *5. DARPINs to neutralise ARBMs of S protein*

Designed ankyrin repeat proteins (DARPins) are small engineered non-immunoglobulin AR proteins (14-21 kDa) (Stumpp et al., 2008), usually composed of four to six AR motifs. They are considered as potential alternatives to monoclonal antibodies (mAbs) (Caputi and Navarra, 2020). DARPins are stable molecules with great affinity (picomolar), specificity and tissue penetration (Plückthun, 2015). In addition, they can be administrated by different routes (oral, nasal, inhaled and topical). This is optimal to directly deliver of high dose of DARPins into oronasal cavity (where SARS-CoV-2 infection starts) and respiratory airways for both prophylactic and therapeutic protection in the early steps of the infection. For example, it has been reported that the entry of human immunodeficiency virus (HIV) into the host cell is blocked by DARPins. The latter compete with the HIV protein for the CD4 binding site on lymphocytes (Tomlinson et al., 2004).

## *6. The respiratory pathophysiology in COVID-19 and role of TRPV1 and TRPV4 in lungs injuries*

To defend the respiratory airways, TRPs induce airway constriction, sneezing and coughing, inflammation, and mucus secretion (Xia et al., 2018; Emir et al., 2017). TRPV4 is expressed in epithelia of the trachea and lungs (Lorenzo et al., 2008), and has an important role in lung and vascular physiology (Rosenbaum et al., 2020). TRPV4 has been proposed as a possible therapeutic target for the treatment of some pulmonary diseases (Rosenbaum et al., 2020). It was demonstrated in animal models of ventilator induced pulmonary injury that TRPV4 played a crucial role in the injury to the lungs (Hamanaka et al., 2007; Rosenbaum et al., 2020). Interestingly, it has been reported that the respiratory pathophysiology in COVID-19 is similar to lungs injuries produced by the sensitization of TRPV1 (Nahama et al., 2020) and TRPV4 (Kuebler et al., 2020). Probably, using TRPV1 and TRPV4 antagonists could inhibit or mitigate inflammation.

## *7. Does the interaction of S protein with TRP channels participate in viral cell entry?*

It has been suggested that other proteins such as neuropilin-1, PIKfyve kinase, integrins, or CD4 (Davanzo et al., 2020) may participate in SARS-CoV-2 cell entry (Cantuti-Castelvetri et al., 2020; Kang et al., 2020; Sigrist et al., 2020). For example, in Hepatitis C virus (HCV), the entry of virus in cell is a complex process that involves five cell surface molecules in binding step, whereas post binding step requires the interaction of HCV nonstructural 5A protein with ARD of ANKRD1 and the internalization of HCV via clathrin-mediated endocytosis (Scheel et al., 2013; Than et al., 2016). Interestingly, it has been reported that the ARDs of TRPV1 and TRPV4 bound to lipids in membrane (Takahashi et al., 2014). Thus, if functional, S-ARBMs could be involved in post binding step by interacting with TRPs ARD which mediate the potential viral entry route to the endocytic-lysosomal pathway, as has been demonstrated in HCV nonstructural 5A protein (Than et al., 2016). Moreover, Sigrist et al. (2020) reported that S protein has an RGD motif that could play a role of co-receptor by interacting with cell surface integrins. Interestingly, the RGD motif is close to $^{408}$RQIAPG$^{413}$ in a hot disordered loop ($^{404}$**RGD**EV**RQIAPG**QTGKIA$^{419}$), suggesting that it may participate in the internalization of S1 subunit.

Different TRP channels are linked to viral infection (Table 3). Thus, respiratory viruses like respiratory syncytial virus (RSV) and measles virus (MV) may interact directly and/or indirectly with TRPA1/TRPV1 on sensory nerves and epithelial cells in the airways (Omar et al., 20017; Harford et al., 2018). Furthermore, rhinovirus can infect neuronal cells and causes upregulation of TRPA1/TRPV1 (Abdullah et al., 2014). Recently, it has been reported that herpes simplex virus 1 (HSV-1) glycoprotein D interacts with TRPC1, and this interaction facilitated the cell entry of HSV-1 (He et al., 2020).

TRPs are principally $Ca^{2+}$ channels and it is known that $Ca^{2+}$ entry in cell plays a major role in infection by several types of virus, such as Sindbis virus, West Nile (Scherbik et al., 2010), HIV, filovirus, and arenavirus (Yao et al., 2012; Han et al., 2015). In addition, it has been demonstrated that TRPV1 regulates $Ca^{2+}$ influx during Chikungunya virus (single positive-stranded RNA virus) infection (Kumar et al., 2020). Recently, it has been reported that the fusogenic ability of MERS-CoV has been improved by over a two-fold increase in intracellular $Ca^{2+}$ (Straus et al., 2020). The same result had already been obtained with SARS-CoV (Madu et al., 2009).

## *8. TRPA1 and TRPV1 gene variants could explain ethnic difference of OTCD?*

It has been reported that Caucasians had a 3-6 times higher prevalence of OTCD than East Asians (von Bartheld et al., 2020). This ethnic difference points to genetic variants in SARS-CoV-2 binding entry proteins in neurons in the nasal cavity and mouth (von Bartheld et al., 2020). TRPA1 is implicated in perception of odors with a trigeminal component (Jordt et al., 2004; Richards et al., 2010). A genetic variant of TRPA1 was associated with enhanced of the sensitivity to odorous stimuli (Schütz et al., 2014), and another genetic variant of TRPA1 was linked to personal differences in the taste perception of cilantro (Knaapila et al., 2012). Furthermore, in humans, a genetic variation in TRPV1 plays a role in salty taste perception (Dias et al., 2013; Chamoun et al., 2018) and the burning sensation from sampled ethanol (Allen et al., 2014).

## *9. SARS-CoV-2, TRPA1/TRPV1 and nicotine?*

Despite of current controversies about the potential protective role of the nicotine in COVID-19 patients (Farsalinos et al., 2020a; Farsalinos et al., 2020b; Richardson et al., 2020; Rossato et al., 2020; Li Volti et al., 2020), I think that it is important to discuss about this relationship because this manuscript suggests that ARBMs of S protein could interact with TRPA1/TRPV1. The latter are expressed in human airway epithelial cells and are in contact with cigarette vapors. TRPA1 mediates the effects of nicotine, and its inhibition with A967079 restored nicotine-mediated impairment of mucociliary function (Chung et al., 2019). In addition, it has been observed that nicotine had a bimodal action on TRPA1, with activation and inhibition occurring at low and high concentrations, respectively (Talavera et al., 2009). Furthermore, nicotine inhibits TRPV1 and elicits taste and smell sensations (Talavera et al., 2009). Probably, the supposed protective effect of nicotine against COVID-19 may be due to the competition between S-ARBMs and nicotine for TRPA1/TRPV1 binding.

## *10. Immune response to SARS-CoV-2 and TRPs?*

The sensory neurons (nociceptors) and immune system work together to defend the organism against external assaults (Tynan et al., 2019; Foster et al., 2017). Indeed, the crosstalk between TRPV1 positive nerve fibers and immune cells is very important in inducing inflammation of the airways after interaction with inhaled allergens or viral particles (Tränkner et al., 2014; Talbot et al., 2015). Moreover, TRPA1 is co-expressed with TRPV1 in nociceptors (Story et al., 2003; Hata et al., 2012) and participates in inflammatory reflex that purpose to mitigate inflammation (Silverman et al., 2019). It has been reported that dysregulated TRPV1 and TRPV4 function has been implicated in lung inflammation (Helyes et al., 2007), and SARS-CoV-2 induces sustained host inflammation (Merad and Martin 2020). This suggests that the sustained inflammation may be the result of the interaction of S protein with ARDs of TRPV1 and/or TRPV4. Some studies showed that SARS-CoV-2 patients developed little antibodies and their persistence is short (Robbiani et al., 2020; Seow et al., 2020; Long et al., 2020). In addition, it has been shown that TRPV1 is required for competent antibody responses to novel antigen (Tynan et al., 2019). This suggests that SARS-CoV-2 weak immunity could be explained by the dysregulation of TRPV1 function by S protein. Probably, using agonists of TRPV1 as adjuvant in SARS-CoV-2 vaccination could enhance the quality and durability of immune response.

## MATERIAL AND METHODS

### *Sequence analysis*

To search probable short linear motifs (SLiMs), SARS-CoV-2 spike protein sequence was scanned with the eukaryotic linear motif (ELM) resource (http://elm.eu.org/). The identified R-x-x-[PGAV][DEIP]-G ARBM was also searched in proteins of SARS-CoV-2, SARS-CoV and MERS-CoV using https://prosite.expasy.org/scanprosite/. All proteins sequences were downloaded from NCBI and UniProt proteins databases.

### *S proteins alignment and phylogeny*

Amino acid residues sequences of SARS-CoV-2 S protein and representative betacoronaviruses (betaCoVs) were aligned with Clustal omega (Sievers et al., 2011) to show if SARS-CoV-2 $^{408}$RQIAPG$^{413}$ and $^{905}$RFNGIG$^{910}$ motifs are conserved in these betaCoVs. This alignment is also used to establish the phylogenetic relationships between these betaCoVs S proteins by constructing a phylogenetic tree with MrBayes (Huelsenbeck and Ronquist, 2001) using: Likelihood model (Number of substitution types: 6(GTR); Substitution model: Poisson; Rates variation across sites: Invariable + gamma); Markov Chain Monte Carlo parameters (Number of generations: 100 000; Sample a tree every: 1000 generations) and Discard first 500 trees sampled (burnin).

*Drugs mimicking R-x-x-[PGAV][DEIP]-G motif*

In the aim to find small molecules containing substituents that topologically and structurally mimic R-x-x-[PGAV][DEIP]-G motif, sketch of 3D structure of $^{408}$RQIAPG$^{413}$ and $^{905}$RFNGIG$^{910}$ are generated and compared with molecules in PubChem (https://pubchem.ncbi.nlm.nih.gov/).

*3D modeling of hTRPA1 and blind docking*

3D structure of AR region of hTRPA1 was modeled using as template the structure of human ankyrin-2 (PDB id: 4rlv_A). The obtained 3D model quality was assessed by analysis of a Ramachandran plot through PROCHECK (Vaguine et al., 1999). For hTRPV1 and hTRPV4, their structure of AR region was obtained from Protein Data Bank (PDB id: 6l93_A) and (PDB id: 4dx1_A), respectively.

Blind docking was conducted by Frodock (Garzon et al., 2009) and AutoDock vina (Trott and Olson, 2010) softwares. To validate the accuracy of the docking by these softwares, subunits of crystal structure of the complex ACE2-RBD (PDB id: 6m0j) and ARD-human peptide SH3BP2 (PDB id: 3twr) were re-docked. Thus, in docking protocol, the coordinates of each separated molecules were used as ligand (6m0j_E) and receptor (6m0j_A) for ACE2-RBD. And ligand (3twr_G) and receptor (3twr_C) for ARD-human peptide SH3BP2 (LPHLQ**RSPPDG**QSFRS). Indeed, Frodock and Autodock Vina were able to produce a similar docking pose for each control protein with respect to its biological conformation in the co-crystallised protein-protein complex.

To test the potential interactions of S protein with AR region of TRPA1 and TRPV1, the 3D coordinates of RBD and HR1 containing $^{408}$RQIAPG$^{413}$ and $^{905}$RFNGIG$^{910}$, respectively, were extracted from structure of SARS-CoV-2 S protein (PDB id: 6vxx_A) and docked with Frodock into AR region of 3D modeled hTRPA1 and hTRPV1. In addition, 3D coordinates of $^{408}$RQIAPG$^{413}$ and $^{905}$RFNGIG$^{910}$ peptides were extracted from the structure of S protein (PDB id: 6vxx_A) and docked using Frodock and AutoDock Vina. The obtained 3D complexes of $^{408}$RQIAPG$^{413}$ and $^{905}$RFNGIG$^{910}$ peptides were refined by using FlexPepDock (London et al., 2011), which allows full flexibility to the peptide and side-chain flexibility to the receptor. The electrostatic potential surface of hTRPA1 and hTRPV1 AR region was realized with PyMOL software (http://pymol.org/).

## ACKNOWLEDGMENTS

I would like to thank the IBIS bioinformatics group for their help.

## CONFLICT OF INTERESTED

The author declares that he has no conflicts of interest.

**FIGURES LEGEND**

**Figure 1**. Multiple amino acid alignment of S proteins of betacoronaviruses (betaCoVs) using Clustal omega and phylogenetic tree. (**A**) $^{408}$RQIAPG$^{413}$ and (**B**) $^{905}$RFNGIG$^{910}$ motifs are indicated by green stars. The figure was prepared with ESPript (http://espript.ibcp.fr). (**C**) Unrooted phylogenetic tree of S proteins of representative betaCoVs. The tree was constructed using Mr Bayes method based on the multiple amino acid sequence alignment by Clustal omega. Red rectangle clusters betaCoVs with both $^{408}$RQIAPG$^{413}$ and $^{905}$RFNGIG$^{910}$ motifs. GenBank and UniProt accession numbers are indicated at the start of each sequence.

**Figure 2**. Localization of ankyrin repeat binding motifs (ARBMs) $^{408}$RQIAPG$^{413}$ and $^{905}$RFNGIG$^{910}$ in S protein of SARS-CoV-2. (**A**) Diagram representation of S protein colored by domain. N-terminal domain (NTD), receptor-binding domain (RBD), subdomains 1 and 2 (SD1-2, orange), protease cleavage site (S1/S2 and S2′), site Fusion peptide (FP), heptad repeat 1 and 2 (HR1 and HR2), central helix (CH), connector domain (CD), transmembrane domain (TM), cytoplasmic tail (CT), and the localization of $^{408}$RQIAPG$^{413}$ in RBD and $^{905}$RFNGIG$^{910}$ in HR1. (**B**) Surface structure representation of the S protein (PDB id: 6VXX_A). $^{408}$RQIAPG$^{413}$ and $^{905}$RFNGIG$^{910}$ are localized in the surface of S protein (blue).

**Figure 3**. Complexes obtained by blind docking of 3D modeled hTRPA1 Ankyrin repeat (AR) region with SARS-CoV-2 RBD (receptor binding domain), HR1 (heptad repeat 1), and $^{408}$RQIAPG$^{413}$ and $^{905}$RFNGIG$^{910}$ peptides. (**A**) RBD into hTRPA1 AR region. (**B**) HR1 into hTRPA1 AR region. (**C**) Complexes of hTRPA1 AR region and $^{408}$RQIAPG$^{413}$ and $^{905}$RFNGIG$^{910}$ peptides which are superposed to AR of Tankyrase-2-human SH3BP2 peptide complex (PDB id: 3twr). (**D**) Electrostatic potential surface representation of the ARD region of 3D modeled hTRPA1 with docked peptides $^{408}$RQIAPG$^{413}$ and $^{905}$RFNGIG$^{910}$. The electrostatic potential surface of modeled hTRPA1 AR region was realized with PyMOL software (http://pymol.org/). The electrostatic potential with negative charge shown in red and positive charge in blue.

**Figure 4**. Complexes obtained by blind docking of hTRPV1 (PDB id: 6l93_A) Ankyrin repeat (AR) region with SARS-CoV-2 RBD (receptor binding domain), HR1 (heptad repeat 1), and $^{408}$RQIAPG$^{413}$ and $^{905}$RFNGIG$^{910}$ peptides. (**A**) RBD into hTRPV1 AR region. (**B**) HR1 into hTRPV1 AR region. (**C**) Electrostatic potential surface representation of the AR region of hTRPV1 with docked peptides $^{408}$RQIAPG$^{413}$ and $^{905}$RFNGIG$^{910}$. The electrostatic potential surface of hTRPV1 AR region was realized with PyMOL software (http://pymol.org/). The electrostatic potential with negative charge shown in red and positive charge in blue.

**Table. 1. ELM motifs of hot disordered loop in SARS-CoV-2 S protein.**

| Elm Name | Instances | Positions | Elm Description | Cell compartment | Pattern |
|---|---|---|---|---|---|
| LIG_RGD | RGD | 403-405 | The RGD motif is recognized by different members of the integrin family | extracellular, integrin | RGD |
| DOC_ANK_TNKS | V**RQIAPGQ** | 407-414 | The Tankyrase binding motif interacts with the ankyrin repeat domain region in Tankyrase-1,2 | nucleus, cytosol | .R..[PGAV][DEIP]G. |
| LIG_FHA_1 | GQTGKIA | 413-419 | Phosphothreonine motif binding a subset of FHA domains | nucleus | ..(T)..[ILV]. |
| MOD_PIKK_1 | PGQTGKI | 412-418 | (ST)Q motif which is phosphorylated by PIKK family member | nucleus, cytosol | ...([ST])Q.. |

**Table. 2. TRPs-ARDs channels locus and taste when it has been reported.**

| TRPs | Locus | Taste | Reference |
|---|---|---|---|
| TRPA1 | Olfactory epithelium | | Kamiyama et al., 2006 |
| TRPC1 | Olfactory epithelium | | Kamiyama et al., 2006 |
| TRPC6 | Olfactory epithelium | | Kamiyama et al., 2006 |
| TRPV1 | Tongue, olfactory epithelium | | Dhakal and Lee, 2019; Nakashimo et al., 2010; Khalifa Ahmed et al., 2009 |
| TRPV1t | Taste buds | Non-specific salt taste | Lyall et al., 2004 |
| TRPV2 | Olfactory epithelium | | Nakashimo et al., 2010; Khalifa Ahmed et al., 2009; Kamiyama et al., 2006 |
| TRPV3 | Tongue, olfactory epithelium | | Dhakal and Lee, 2019; Nakashimo et al., 2010 Khalifa Ahmed et al., 2009 |
| TRPV4 | Taste buds, olfactory epithelium | | Doñate-Macián et al., 2018 |
| TRPV4 | Mouse gastrointestinal tract | | Matsumoto et al., 2019 |
| TRPV4 | Type IV of the taste buds | Sour | Matsumoto et al., 2019 |
| TRPV4 | Epidermal keratinocytes | | Chen et al., 2014 |
| TRPV4 | Olfactory and airway epithelia | | Ueda et al., 2015 |
| TRPV6 | Olfactory epithelium | | Kamiyama et al., 2006 |

**Table. 3. TRPs-ARDs channels that are linked to viral infection.**

| TRPs | Viruses | Reference |
|---|---|---|
| TRPA1, TRPV1 | Respiratory syncytial virus (RSV) and measles virus (MV) | Omar et al., 2017 |
| TRPC1 | HSV-1 glycoprotein D facilitating entry of virus | He et al., 2020 |
| TRPV1 | Regulatory role during chikungunya virus (CHIKV) infection in macrophages | Sanjai Kumar et al., 2020 |
| TRPV1 | Human rhinovirus (HRV) | Abdullah et al., 2014 |
| TRPV1 | Herpes simplex virus type 1, 2 (HSV1, 2) | Cabrera et al., 2016 |
| TRPV1 | Hepatitis C virus (HCV) | Alhmada et al., 2017 |
| TRPV1 | Varicella-zoster virus (VZV) | Han et al., 2016 |
| TRPV4 | Mediates infectivity of dengue, hepatitis C and Zika Viruses (RNA viruses) | Doñate-Macián et al., 2018 |

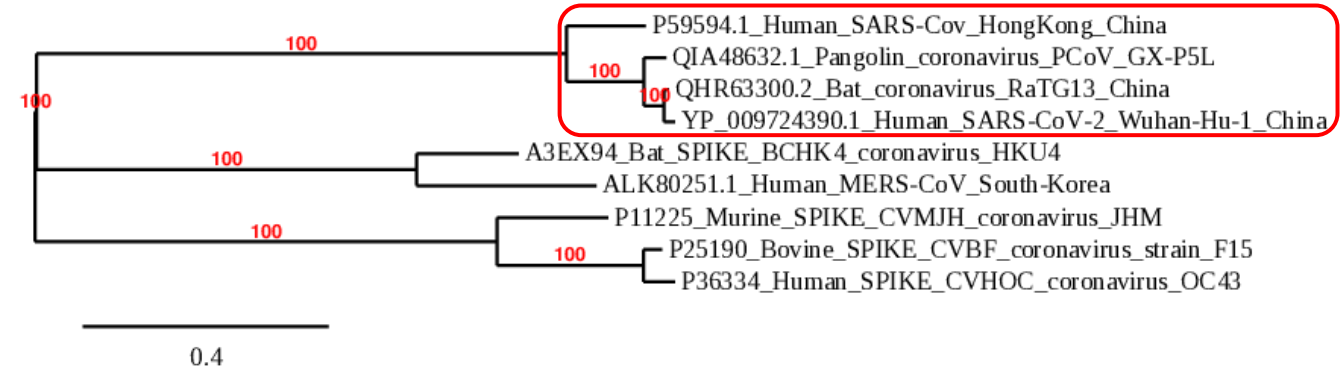

**Fig. 1**

A

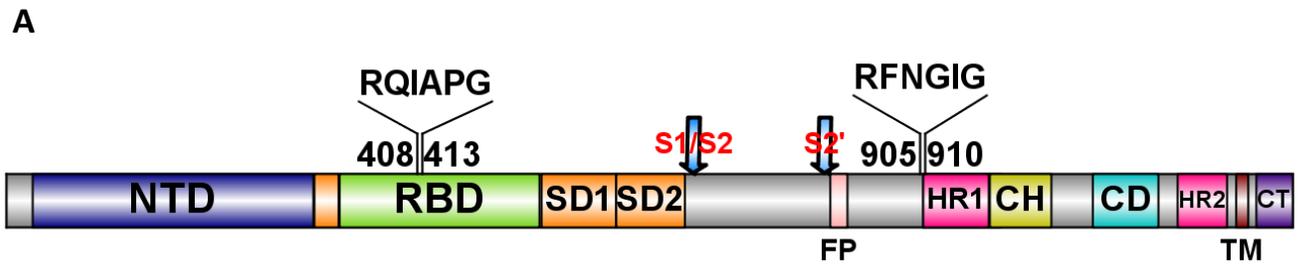

B

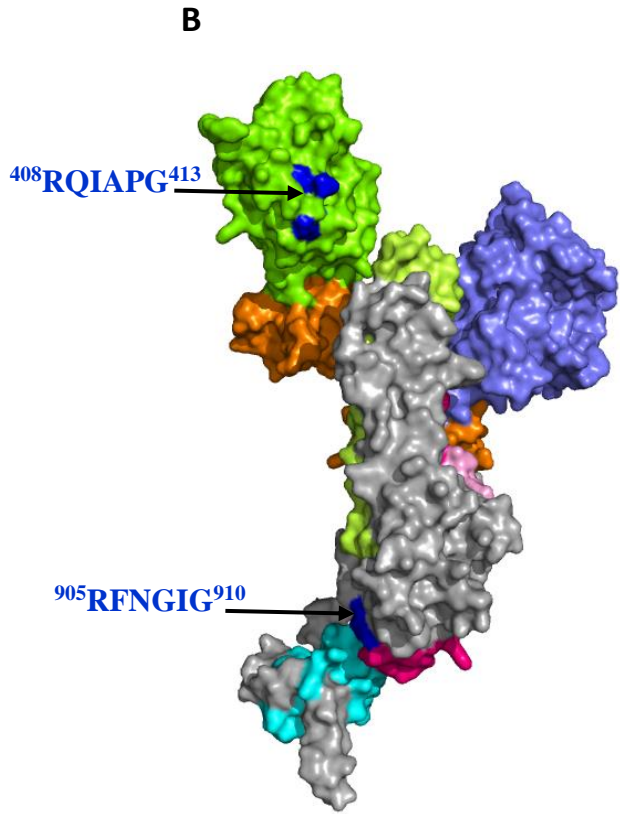

**Fig. 2**

A

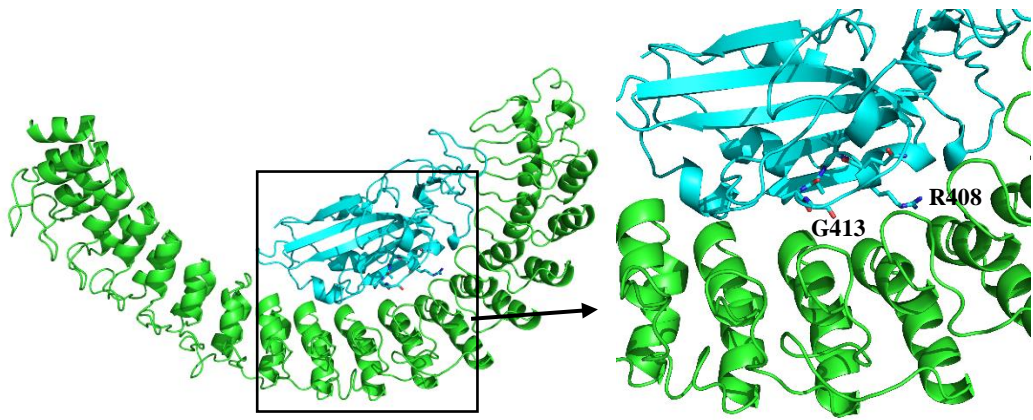

B

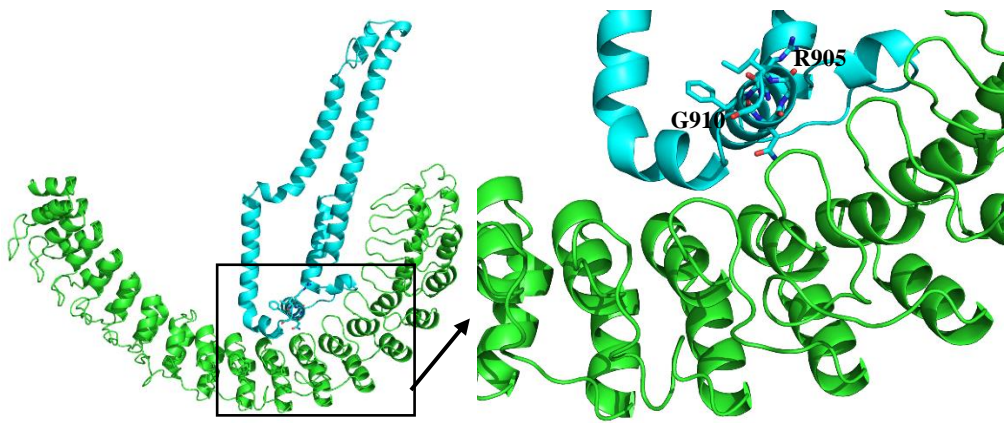

C

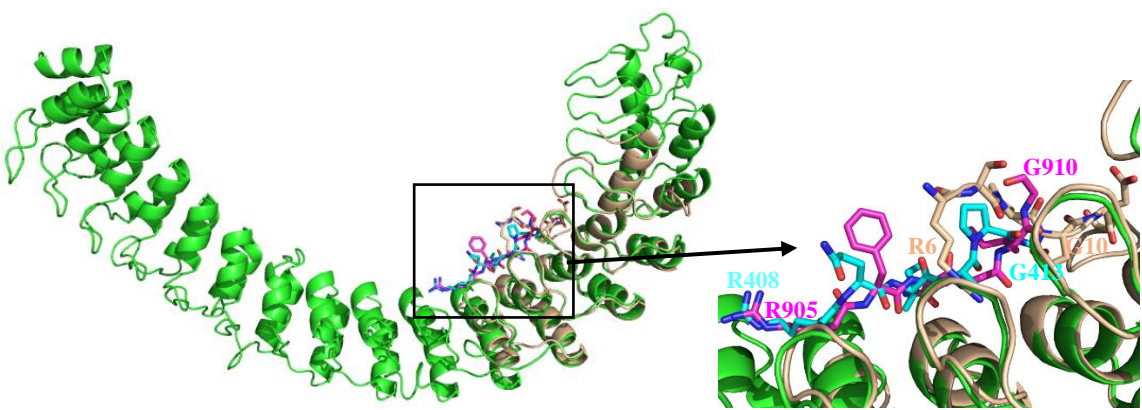

**D**

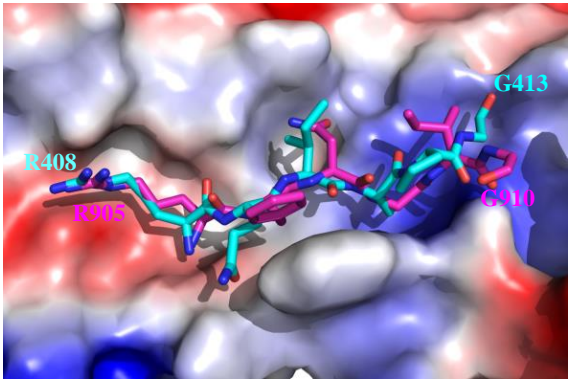

**Fig. 3**

**A**

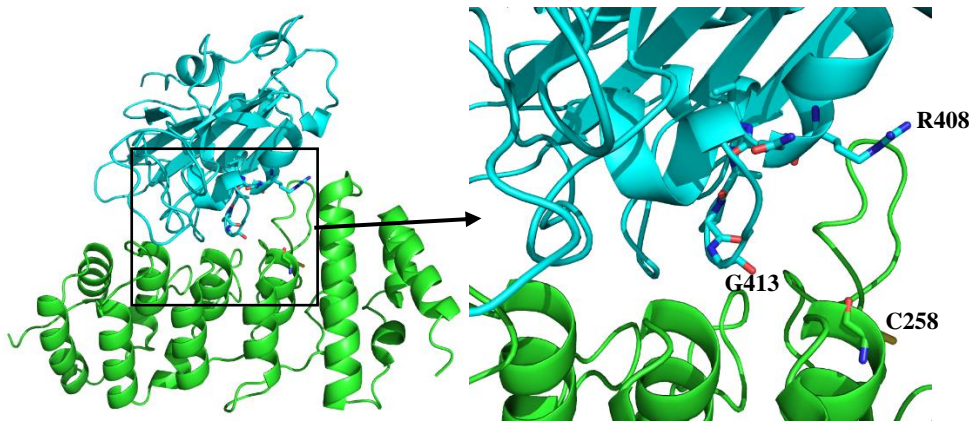

**B**

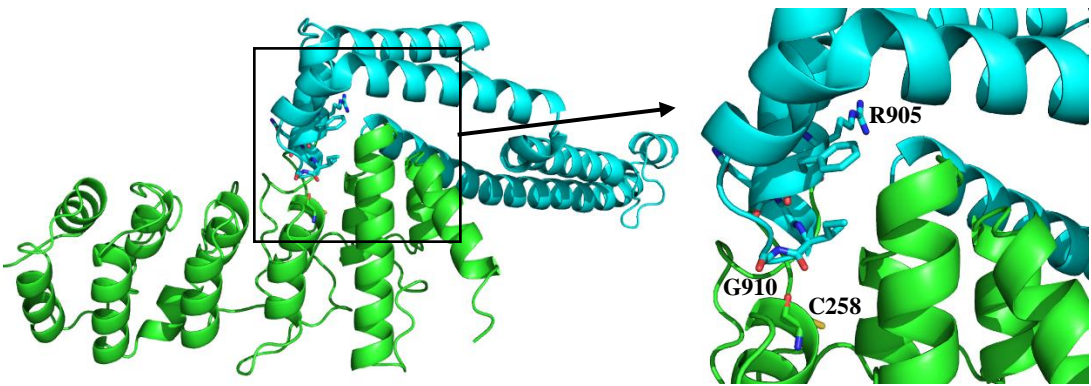

**C**

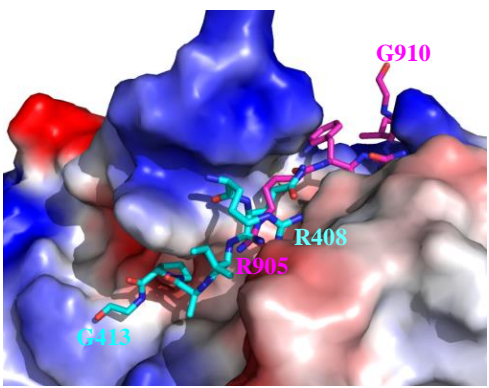

**Fig. 4**